\documentclass[12pt,letterpaper]{iopart}
\usepackage{amsmath}
\usepackage{iopams}
\usepackage{graphicx}
\usepackage{amssymb}

\newcommand{\ket}[1]{|{#1}\rangle  }

\newcommand{\ketbra}[2]{\vert {#1} \rangle \langle{#2}\vert}
\providecommand{\openone}{\leavevmode\hbox{\small1\kern-3.8pt\normalsize1}}

\begin{document}

\title{Hidden entanglement,      
system-environment information flow and non-Markovianity}

\author{A. D'Arrigo$^{a,b,c}$, G. Benenti$^{d,e}$, R. Lo Franco$^{f}$, 
        G. Falci$^{b,a,g}$ and E. Paladino$^{b,a,g}$}
\address{$^a$CNR-IMM UOS Universit\`a (MATIS), Consiglio Nazionale delle Ricerche,
         Via Santa Sofia 64, 95123 Catania, Italy}
\address{$^b$Dipartimento di Fisica e Astronomia, Universit\`a degli Studi Catania, 
         Via Santa Sofia 64, 95123 Catania, Italy}
\address{$^c$Centro Siciliano di Fisica Nucleare e Struttura della Materia (CSFNSM),
            Via Santa Sofia 64, 95123 Catania, Italy}
\address{$^d$CNISM and Center for Nonlinear and Complex Systems,
            Universit\`a degli Studi dell'Insubria, Via Valleggio 11, 22100 Como, Italy}
\address{$^e$Istituto Nazionale di Fisica Nucleare, Sezione di Milano,
             via Celoria 16, 20133 Milano, Italy}
\address{$^f$Dipartimento di Fisica e Chimica, Universit\`a di Palermo, 
             via Archirafi 36, 90123 Palermo, Italy}
\address{$^g$Istituto Nazionale di Fisica Nucleare, Sezione di Catania,
            Via Santa Sofia 64, 95123 Catania, Italy}

\begin{abstract}
It is known that entanglement dynamics of two noninteracting qubits,
locally subjected to classical environments, may exhibit revivals.
A simple explanation of this phenomenon may be provided by using the concept of
hidden entanglement, which signals the presence of entanglement that
may be recovered without the help of nonlocal operations.
Here we discuss the link between hidden entanglement and
the (non-Markovian) flow of classical information between the system 
and the environment.
\end{abstract}


\maketitle

\section{Introduction}	

A bipartite system $\textsf{S}$ can exhibit entanglement revivals 
even in the case in which the two subsystems are noninteracting and affected by 
local independent classical noise sources~\cite{lofrancorandom,darrigo2012}.
That is to say, entanglement can be a non-monotonic function of time,
increasing in some time interval in spite of the lack of non-local operations
or entanglement back transfer from a quantum environment~\cite{bellomo2007PRL}.  
This phenomenon is at first sight surprising, since we consider 
conditions under which entanglement, which is a non-local resource, cannot
be created. Therefore the increase of entanglement must be attributed to 
the manifestation of quantum correlations that were already present
in the system.
The density operator formalism does not capture the presence of these quantum 
correlations, thus they are in some sense \textit{hidden}.
On the other hand, as we showed in Ref.~\cite{darrigo2012}, 
the existence of such correlations is enlightened 
if the system is described as a physical ensemble of states.
For this reason we introduced the concept of 
{\em hidden entanglement} (HE)~\cite{darrigo2012}.

The hidden entanglement $E_h(t)$ is the amount of entanglement that 
cannot be exploited at a given time $t$, due to a lack of classical 
information on the system $\textsf{S}$. 
As a result, at times $\tilde{t}>t$ this entanglement may be recovered 
without the help of any non-local operation. 
In other words, classical communication and/or local operations can succeed in 
increasing entanglement because they are not required to create any entanglement,
which is impossible. They only remove that lack of knowledge which at time $t$ 
forbids us to dispose of the hidden amount of entanglement given by $E_h(t)$.

Relevant examples where the environment can be modeled as a classical 
system may be found in solid-state implementations of qubits, with
the dynamics of such nanodevices dominated by low-frequency 
noise~\cite{FalciPrl05,Ithier2005,Bylander,Chiarello,revivalstochastic,PaladinoRMP2013}.
The recovery of quantum correlations in a classical environment
was demonstrated in an all-optical experiment~\cite{cenv_nature}.

In this paper, we discuss the relationship between recovery of
HE and information flow from a classical environment to the system,
quantified by a decrease in time of their quantum mutual information. 
We show that such non-Markovian behavior by itself does not 
guarantee entanglement recovery. It is also necessary that the 
ensemble physically underlying the system's mixed state
includes entangled states, whose entanglement content can be 
recovered without nonlocal operations.  

\section{Hidden entanglement}
\label{sec:HE}
Let us start by considering a bipartite system $\textsf{S}$ whose dynamics is 
described by an ensemble of states ${\cal A}(t)=\{(p_i(t),|\psi_i(t)\rangle)\}$.  
That is, we know the statistical distribution of the bipartite pure states
$\{|\psi_i(t)\rangle\}$, occurring with probabilities
$\{p_i(t)\}$, so that 
$\rho(t)=\sum_i p_i(t) |\psi_i(t)\rangle\langle \psi_i(t)|$,
but the state of any individual system in the
ensemble is unknown. 
The average entanglement of the ensemble ${\cal A}(t)$ is defined 
as~\cite{Bennett96,Cohen98,Nhal04,Carvalho07}:
\begin{equation} 
   {E}_{av}({\cal A},t)=\sum_i p_i(t) E(|\psi_i(t)\rangle\langle \psi_i(t)|),
\label{eq:averageEntanglement}
\end{equation}
where $E(|\psi\rangle\langle \psi|)=
S(\rho_\textsf{A})=S(\rho_\textsf{B})$ is the 
entropy of entanglement, with 
$\rho_\textsf{A}={\rm Tr}_\textsf{B} (|\psi\rangle\langle \psi|)$ 
and $\rho_\textsf{B}={\rm Tr}_\textsf{A} (|\psi\rangle\langle \psi|)$ 
reduced states of subsystems $\textsf{A}$ and $\textsf{B}$.
On the other hand, the 
entanglement\cite{PlenioReview,HorodeckiReview} 
of the state $\rho(t)$ is quantified by some convex entanglement measurement 
$E(\rho(t))$ (reducing to the entropy of entanglement for pure states).
We defined in Ref.~\cite{darrigo2012} the {\it hidden entanglement} 
of the ensemble ${\cal A}(t)=\{(p_i(t),|\psi_i(t)\rangle)\}$ as
the difference between the average entanglement of the ensemble ${\cal A}(t)$,
and the entanglement of the corresponding state $\rho(t)$
\footnote{Note that in Ref.~\cite{Cohen98} the expression
``hidden entanglement'' is used with a different meaning.}:
\begin{equation}
\begin{array}{l}
{\displaystyle
    {E_h}({\cal A},t) \equiv E_{av}({\cal A},t) - E(\rho(t))\,=
}
\\
{\displaystyle\hspace{0.5cm}
= \sum_i p_i (t) E(|\psi_i(t)\rangle\langle\psi_i(t)|)-
E\left(\sum_i p_i(t) |\psi_i(t)\rangle\langle\psi_i(t)|\right).
}
    \label{eq:MeasHiddenEntang}
\end{array}
\end{equation} 
Due to the convexity, 
$E_h$ is always larger than or equal to zero.
Several physical examples illustrating the concept of hidden entanglement
are discussed in Refs.~\cite{darrigo2012,hiddenproceeding}.

The HE quantifies the entanglement 
that cannot be exploited as a resource due to the
\emph{lack of knowledge} about which state of the mixture 
we are dealing with. Such entanglement can be recovered
(unlocked~\cite{Cohen98,Eisert00,Gour07}), once such classical information is
provided, \textit{without the help of any non local operation}.

From the definition Eq.~(\ref{eq:MeasHiddenEntang}), it is clear 
that HE is associated with the specific quantum ensemble description of the system 
state. We will refer to situations where the system dynamics 
\textit{admits a single physical decomposition in terms 
of an ensemble of pure state evolutions}. This is always possible, at least 
in principle, when the  system is affected by 
\textit{classical noise sources}~\cite{darrigo2012}. In the case in which the system
interacts with a quantum environment, a physical description of the system 
in terms of quantum ensemble it is still possible, when the 
environment is subjected to a projective 
measurement~\cite{darrigo2012,Nhal04,Carvalho07,mascarenhas10,mascarenhas11,volgelsberger,barchielli,murch}. 
In this case $ E_{av}$ is obtained after averaging over the measurement records
and the hidden entanglement can be recovered by purely local operations 
acting on subsystems $\textsf{A}$ and $\textsf{B}$, provided the classical record 
comprising information read from the environment is accessible.

Since hidden entanglement recovery requires a back-flow of classical information
 from the environment to the system, it is 
interesting to compare HE with the system-environment quantum mutual information
and to discuss entanglement revivals in the context of non-Markovian 
environments~\cite{breuer}.
For this purpose, we embed the randomness present in the quantum 
ensemble ${\cal A}(t)=\{(p_i(t),|\psi_i(t)\rangle)\}$
into the degrees of freedom of a dummy quantum system 
$\textsf{E}$~\cite{Eisert00,devetak2005}, in the following way: 
\begin{eqnarray}
&&\rho(t)=\rho^\textsf{S}(t)=\textrm{Tr}_\textsf{E}\big[\rho^\textsf{SE}(t)\big],\\
&& \rho^\textsf{SE}(t)=\sum_i p_i(t)\ketbra{x_i(t)}{x_i(t)}
 \otimes |\psi_i(t)\rangle\langle\psi_i(t)|,  
\label{eq:rho_in_EHS}
\end{eqnarray}
where $\{\ket{x_i(t)}\}$ is an orthonormal basis for \textsf{E} at time $t$.  
From (\ref{eq:rho_in_EHS}) one can compute the 
quantum mutual information~\cite{nielsen-chuang,benenti-casati-strini} 
between the fictitious environment $\textsf{E}$ and the system $\textsf{S}$: 
\begin{equation}
I(\textsf{S}:\textsf{E})=S(\rho(t))\,+\,S(\rho^\textsf{E}(t))\,-\,
S(\rho^\textsf{SE}(t))),
\label{eq:mutual-information-1}
\end{equation}
where
$\rho^\textsf{E}(t)=\textrm{Tr}_\textsf{S}\big[\rho^\textsf{SE}(t)\big]$.
Using the relation~\cite{nielsen-chuang}
$S\big(\sum_i p_i \ketbra{x_i}{x_i}\otimes\rho_i\big)=
H(p_i)+\sum_i p_i S(\rho_i)$, where
it is assumed that $\ket{x_i}$ are orthogonal states,
we obtain
\begin{equation}
I(\textsf{S}:\textsf{E})=S(\rho(t))\,+\,H({p_i(t)})\,-\,H({p_i(t)})= S(\rho(t)),
\end{equation}
where $H(p_i)=-\sum_i p_i \log_2p_i$ is the 
Shannon entropy associated with the
probability distribution $\{p_i\}$.
Therefore, the loss of knowledge on the system, and the 
resulting uncertainty on the system state $S(\rho)$, can be interpreted as 
due to a flux of information~\cite{breuer} 
between $\textsf{S}$ and $\textsf{E}$, the total amount of information 
shared between system and environment being quantified by the quantum mutual 
information~\cite{mazzola2012,luo}. 

\section{Entanglement revivals under random local fields}
\begin{figure}[t!]
\centerline{\includegraphics[angle=0.0, width=8cm]{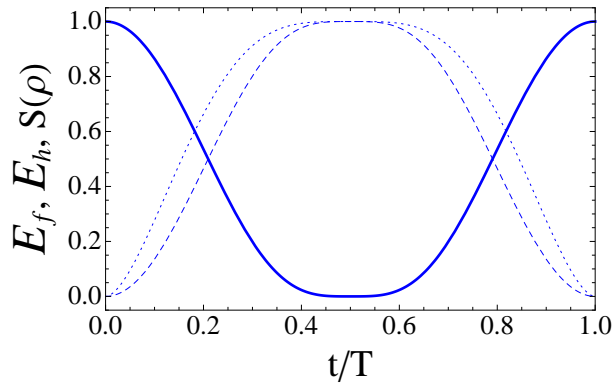}} 
\vspace*{8pt}
\caption{Entanglement of formation (solid curve) 
$E_f(\rho(t))$ for the system described by the
quantum ensemble ${\cal A}(t)$ of Eq.~(\ref{eq:Q-ensemble1}). We also show the 
hidden entanglement $E_h({\cal A},t)$ (dashed curve), and the 
von Neumann entropy (dotted curve) $S(\rho(t))$.}
\label{fig:example-1} 
\end{figure}

We illustrate the similarities and differences between 
HE and quantum mutual information in the example
of entanglement revivals under random local fields 
that we have studied in Ref.~\cite{darrigo2012}.
Let us first consider a two-qubit system $\textsf{AB}$ initially
prepared in the maximally entangled Bell state $|\phi^+\rangle$.
The time evolution consists of local unitaries, but we have no 
complete information about which local unitary is acting. 
In particular, we suppose that
qubit $\textsf{A}$ undergoes, with equal probability, 
a rotation about the $x$-axis of its Bloch sphere, 
$U_x(t)=\mathrm{e}^{-\mathrm{i} \sigma_x \omega t/2}$,
or a rotation around the $z$-axis, 
$U_z(t)=\mathrm{e}^{-\mathrm{i} \sigma_z \omega t/2}$,
while qubit $\textsf{B}$ remains unchanged.
Hence, the ensemble ${\cal A}$ at time $t$ is
\begin{equation}
{\cal A}(t)=\big\{\big(p_i,\mathbb{U}_i(t)|\phi^+\rangle\big)\big\}\,=\,
\Big\{\Big(\frac{1}{2}, \mathbb{U}_x(t) \ket{\phi^+}\Big),
\,\Big(\frac{1}{2}, \mathbb{U}_z(t)  \ket{\phi^+}\Big)\Big\},
\label{eq:Q-ensemble1}
\end{equation}
where $\mathbb{U}_i\equiv U_i(t) \otimes \openone_\textsf{B}$.
Since we are dealing with random local unitaries, the average 
entanglement of ${\cal A}$ is constant in time, $E_{av}({\cal A}(t))=1$.
On the other hand, the entanglement of the corresponding state 
$\rho(t)=\frac{1}{2}(\mathbb{U}_x\ketbra{\phi^+}{\phi^+}\mathbb{U}^\dag_x+
\mathbb{U}_z\ketbra{\phi^+}{\phi^+}\mathbb{U}^\dag_z)$ 
changes in  time, see Fig.~\ref{fig:example-1}
\footnote{Here and in the following, as entanglement measure we use the 
entanglement of formation $E_f(\rho)$, which is an upper bound for any 
bipartite entanglement measure~\cite{horodeckiprl}. Therefore $E_{av}-E_f(\rho)$ 
is a lower bound for HE; moreover
$E_f(\rho)$ can be readily computed for two-qubit systems via 
the concurrence $C(\rho)$~\cite{Wootters98}.}.  
At $\overline{t}=T/2$ ($T\equiv\frac{2\pi}{\omega}$),
$\rho(\overline{t})=\frac{1}{2}|\phi^-\rangle\langle\phi^-|+
\frac{1}{2}|\psi^+\rangle\langle\psi^+|$ is separable, whereas
at $2 \overline{t}$, $U_x(2 \overline{t})=U_z(2 \overline{t})=\openone_A$ and 
the initial maximally entangled state is recovered
(we use the notation $|\psi^\pm\rangle= (\ket{01}\pm\ket{10})/\sqrt{2}$,
$|\phi^\pm\rangle= (\ket{00}\pm\ket{11})/\sqrt{2}$).
In the interval $[\overline{t},2\overline{t}]$ the entanglement revives from
zero to one without the action of any nonlocal quantum operation, 
thus apparently violating the monotonicity axiom.
The ensemble description tells us that at time
$\overline{t}$ the system is always in an entangled state 
($\ket{\phi_-}$ or $\ket{\psi_+}$), but the lack of knowledge about 
which local operation the system underwent prevents us from distilling 
any entanglement: entanglement is {\em hidden}, 
${E_h}({\cal A}(\overline{t}))=1$ and $E_f(\rho(\overline{t}))=0$.
At time $2\overline{t}$ this lack of knowledge
is irrelevant since the two possible time evolutions result
in the identity operation $\openone_\textsf{A}$ and entanglement is recovered, 
${E_h}({\cal A}(2\overline{t}))=0$ and $E_f(\rho(2\overline{t}))=1$.

As we can see from Fig.~\ref{fig:example-1}, the von Neumann entropy 
$S(\rho(t))$, which is equal to the quantum mutual information 
$I(\textsf{S}:\textsf{E})$ since we are considering an ensemble of 
pure states (see Sec.~\ref{sec:HE}),
exhibits a behavior very close to that of HE. 
This similarity has a simple explanation: We have an 
ensemble ${\cal A}(t)$ of maximally entangled states, so that 
the lack of knowledge about which state of the mixture we
are dealing with is the only reason that prevents us from exploiting 
entanglement as a resource. Such lack of knowledge is correctly captured
by the von Neumann entropy.  

Let us now consider the same quantum operation described at the beginning of this
section, but starting from a different initial state, in general mixed:
\begin{equation}
\rho_0=\eta\ketbra{\phi^+}{\phi^+}+
     (1-\eta)\Big(\frac{1}{2}\ketbra{00}{00}+\frac{1}{2}\ketbra{11}{11}\Big).
\label{eq:rho0}
\end{equation}
That is, the system is prepared in the maximally 
entangled state $\ket{\phi_+}$ with probability $\eta$, 
or in one of the separable states $\ket{00}$ and $\ket{11}$, with equal 
probability $\frac{1}{2}(1-\eta)$, and the preparation record is disregarded
or not available. 
In this case the quantum ensemble at time $t$ is
\begin{equation}
{{\cal A}}(t)=\big\{\big(p_i,\mathbb{U}_i(t)\rho_0\mathbb{U}^\dag_i(t)\big)\big\}
\,=\,
\Big\{\Big(\frac{1}{2}, \mathbb{U}_x(t) \rho_0 \mathbb{U}_x^\dag(t)\Big),
\,\Big(\frac{1}{2}, \mathbb{U}_z(t)  \rho_0 \mathbb{U}_z^\dag(t)\Big)\Big\}.
\label{eq:Q-ensemble3}
\end{equation}
The density operator which arises from such ensemble is
$\rho(t)=\sum_i p_i \mathbb{U}_i(t) \rho_0 \mathbb{U}^\dag_i(t)$.
The parameter $\eta$ allows us to set
simultaneously the entanglement and the mixedness of the 
initial state $\rho_0$. 
When $\eta=1$ the system $\textsf{S}$ is initially in a maximally entangled pure 
state and we recover the case discussed at the beginning of this section; 
when $\eta=0$, the initial state of $\textsf{S}$ is a separable mixed state.
We define the hidden entanglement 
associated with the ensemble (\ref{eq:Q-ensemble3}) as
\begin{equation}
E_h({{\cal A}},t)=
\sum_i p_i E(\mathbb{U}_i(t)\rho_0\mathbb{U}_i^\dag(t))- E(\rho(t))=
E_f(\rho_0)- E_f(\rho(t)).
\label{eq:HE-ensemble-tilde2}
\end{equation}
Such definition extends the hidden entanglement measure of 
Eq.~(\ref{eq:MeasHiddenEntang}) to ensemble of mixed states:
The maximum entanglement which we can obtain at any time, once 
we retrieve the information about which operation the system 
undergoes, is just $E_f(\rho_0)$
\footnote{It is worth mentioning that for the above state 
$\rho_0$, $E_f(\rho_0)$ has 
a clear physical meaning, being equal to the entanglement 
cost~\cite{vidal02}.}.

\begin{figure}[t!]
\centerline{\includegraphics[angle=0.0, width=6cm]{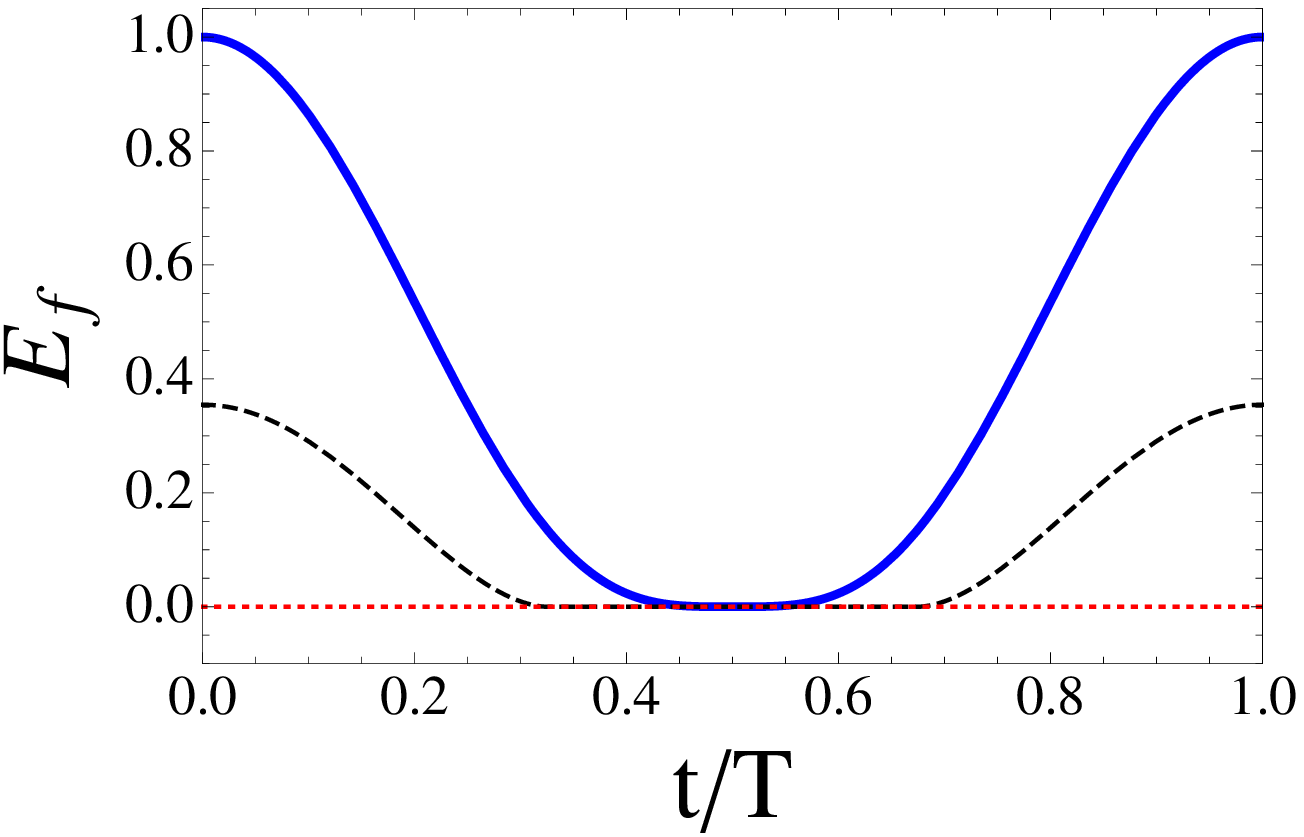}
            \includegraphics[angle=0.0, width=6cm]{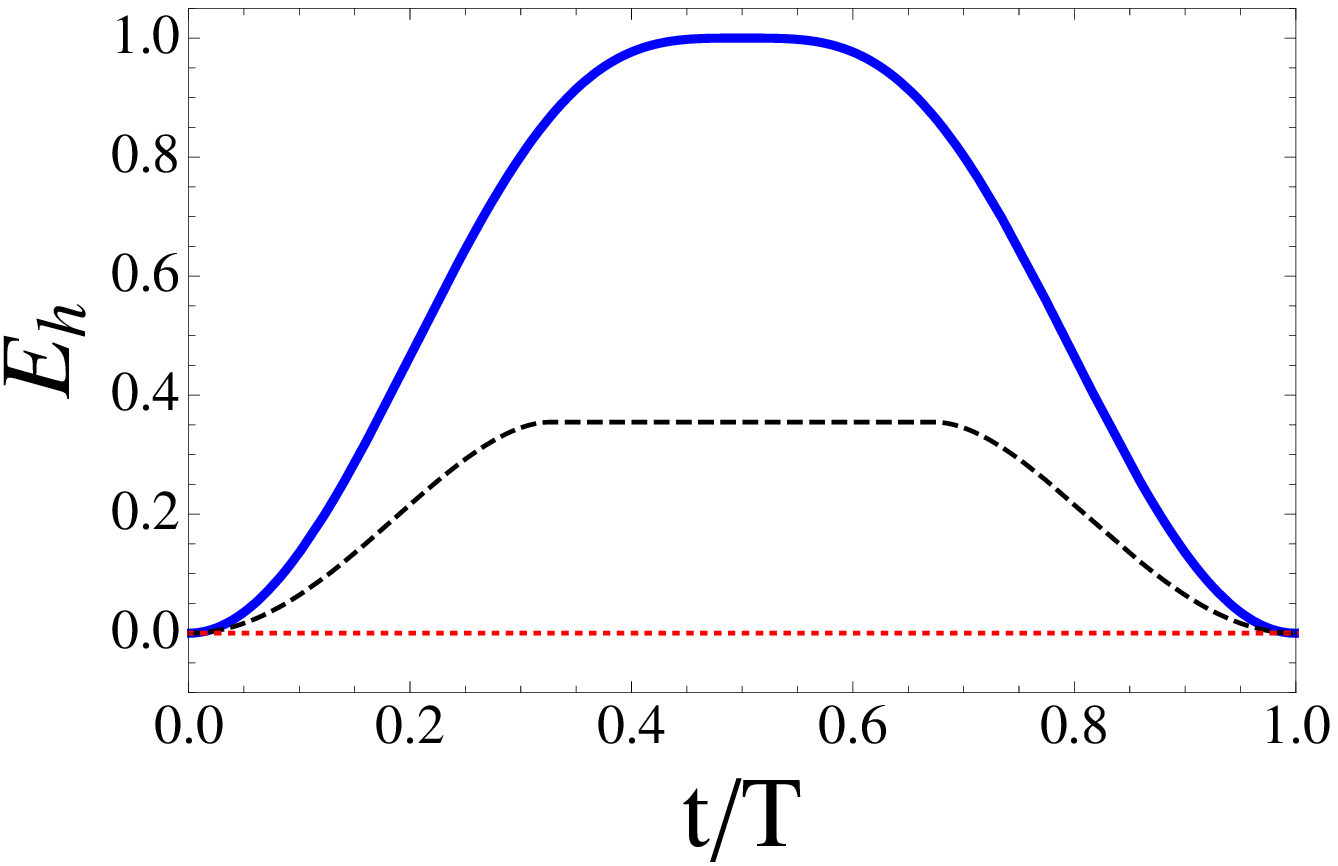}}
\centerline{\includegraphics[angle=0.0, width=6cm]{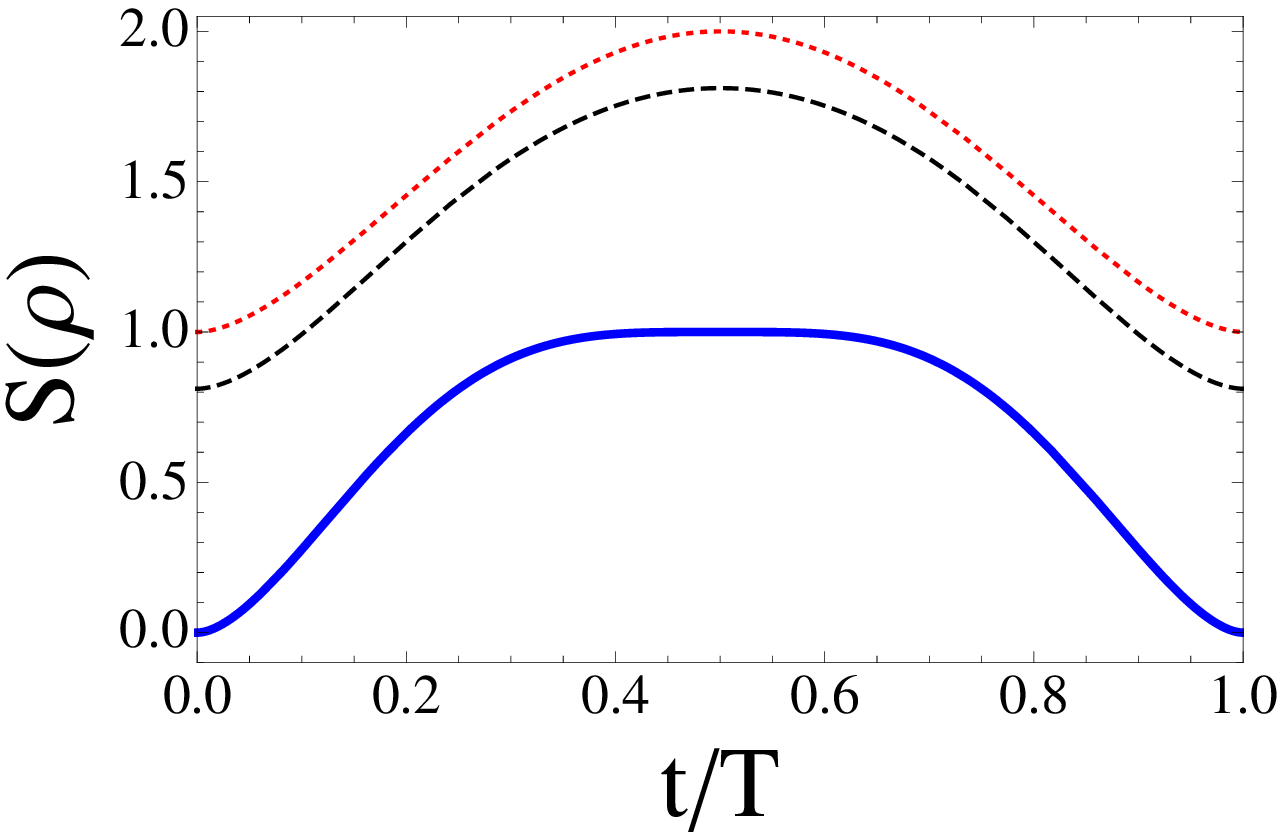}
            \includegraphics[angle=0.0, width=6cm]{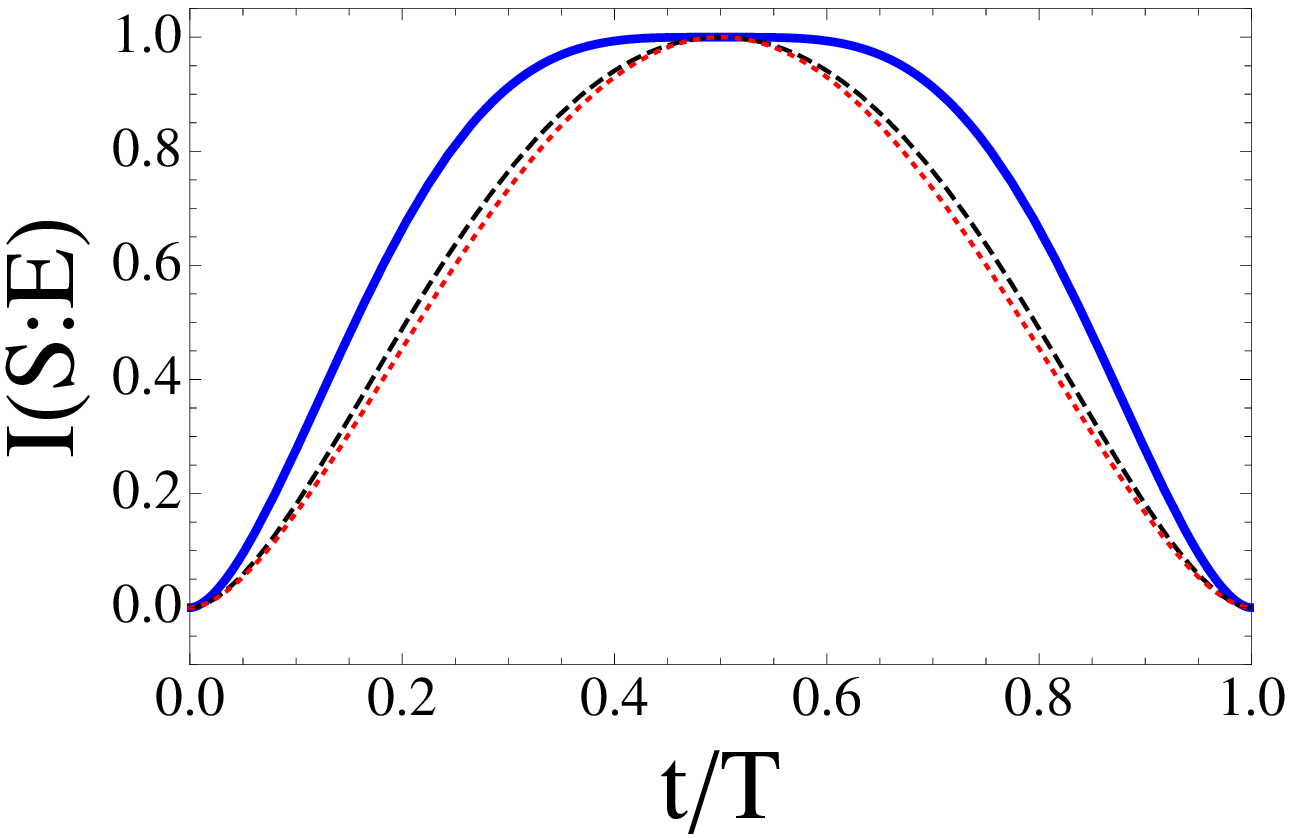}}
\vspace*{8pt}
\caption{Entanglement of formation (top left panel), 
hidden entanglement (top right panel), von Neumann entropy 
(bottom left panel), and system-environment quantum mutual information 
(bottom right panel) 
for the ensemble (\ref{eq:Q-ensemble3}), 
as a function of the time, for 
$\eta=1$ (solid curves),  $0.5$ (dashed curves), and $0$ (dotted curve). 
Data for $\eta=1$ are
the same as in Fig.~\ref{fig:example-1} and are reported here for comparison.}
\label{fig:EntanglementEnsemble2}
\end{figure}

We plot 
in the top panels of Fig.~\ref{fig:EntanglementEnsemble2} 
the entanglement of formation $E_f$ of the state $\rho(t)$
(left) and the hidden 
entanglement  $E_h({{\cal A}},t)$ (right),
for different values of the parameter $\eta$. 
For $\eta=1$ (solid curves) we recover the results of Fig.~\ref{fig:example-1}.
For $\eta=0$ (dotted curves), there is no entanglement in the 
initial state and therefore no entanglement can be generated
during the purely local time evolution, so both $E_f$ and $E_h$ are vanishing 
at any times. For $\eta=0.5$ (dashed curves), first the entanglement 
monotonically decreases until
it sudden dies at $t/T\approx 0.33$ \footnote{Note that while there is 
sudden death of the entanglement, the coherence terms of the density operator
$\rho$ do not vanish.}, 
then it revives after $t/T\approx 0.67$, up to its initial value.
The entanglement decrease and its sudden death are consequences of our ignorance
about the random operation $\mathbb{U}_i$ the system undergoes: 
if this information is provided not only the sudden death may be 
avoided, but also all the initial 
entanglement may be recovered at any time.

In the bottom panels of 
Fig.~\ref{fig:EntanglementEnsemble2} we show the von 
Neumann entropy $S$ of the state $\rho$ (left) and the 
system-environment quantum mutual information $I(\textsf{S}:\textsf{E})$ (right).
For $\eta<1$, $S(\rho(t))$  
exhibits a behavior which is qualitatively and quantitatively different 
from that of  $E_h(\tilde{{\cal A}},t)$. 
For $\eta=0$, $S(\rho(t))$ varies in the range $[1,2]$, 
while the entanglement of formation 
and the hidden entanglement are vanishing at any times.
For intermediate values, for instance $\eta=0.5$, 
$S(\rho(t))$ monotonically increases from the finite 
value $S(\rho_0)$ up to its maximum value reached at $t=T/2$; in the same interval, 
$E_h$ starts from zero and saturates before $T/2$ to its maximum 
value $E_f(\rho_0)$. 
This example shows that the system uncertainty, quantified by the von Neumann
entropy, is not always connected to the possibility of recovering entanglement.
By injecting classical information we reduce 
our lack of knowledge on the system, however such
information does not always allow us to increase the system entanglement.
  
We now consider the quantum mutual information. 
As we have written in Sec.~\ref{sec:HE}, it is possible to introduce a fictitious
environment which simulates the action of our quantum operation.
The system environment state is
\begin{eqnarray}
&& \rho^\textsf{SE}(t)=\sum_i p_i\ketbra{x_i}{x_i}
 \otimes \mathbb{U}_i(t)\rho_0  \mathbb{U}^\dag_i(t).
\label{eq:rho2_in_EHS}
\end{eqnarray}
The quantum mutual information between 
$\textsf{S}$ and $\textsf{E}$ is given by
\begin{eqnarray}
I(\textsf{S}:\textsf{E})&&=S(\rho(t))\,+\,S(\rho^\textsf{E}(t))\,-\,
S(\rho^\textsf{SE}(t)))\nonumber\\
&&=S(\rho(t))\,+\,H({p_i})\,-\,\Big(H({p_i})+\sum_i p_i S(\mathbb{U}_i(t)\rho_0 
\mathbb{U}^\dag_i(t))\Big)= \nonumber\\
&&= \,S(\rho(t))\,-\,S(\rho_0).
\label{eq:mutual_information_2}
\end{eqnarray}
From Fig.~\ref{fig:EntanglementEnsemble2}, we can see
that the quantum mutual information 
is scarcely sensitive to the entanglement that can be recovered.
One can say that $\textsf{S}$ and $\textsf{E}$ develop the same degree
of correlations as a function of the time, regardless of $\eta$, that is independently
of the fact the system is initially prepared in an entangled state or in a 
separable state. In other words, from the value assumed by $I(S:E)$ one cannot
in general estimate the amount of entanglement that is possible to recover
at a given time by classical communication and local operations on the subsystems 
composing $\textsf{S}$.

\section{Final remarks}

The above considerations on the quantum mutual information tell us that
non-Markovianity of system dynamics, which can be interpreted as a
back-flow of information between $\textsf{S}$ and $\textsf{E}$~\cite{breuer} and
quantified~\cite{mazzola2012,luo} in terms of negative time derivative of 
the coherent information, $\partial I(\textsf{S}:\textsf{E})/\partial t\,<\,0$, 
is only a necessary condition for entanglement revivals.
Indeed we can have back-flow of classical information from the 
environment to the system without entanglement revivals, 
provided we are dealing with an ensemble of separable states for the system.
We can conclude with the following statements about the occurrence of entanglement
revivals when a system interacts with classical noise sources.
In the case $E_f(t)=0$, the information we can
acquire from the environment may be useful to recover some entanglement only
if the quantum ensemble physically underlying the system dynamics has a non
vanishing average entanglement. In a more general case, in the presence 
of a back-flow of information and $E_f(t)\neq 0$, we can
identify a sufficient condition for entanglement revivals at time
larger than $t$ by the requirement $E_h(t)>0$.

\ack
This work was partially supported by 
PON02-00355-339123 - ENERGETIC and by 
the MIUR-PRIN project \textit{Collective quantum phenomena:
From strongly correlated systems to quantum simulators}.
E. P. and G. F. acknowledge support from CSFNSM Catania.


\end{document}